\documentclass[onecollarge]{svjour2}       
\usepackage[russian,english]{babel}
\usepackage{graphicx}
\usepackage{amsmath}
\usepackage{cite}
\journalname{}
\begin{document}

\title{Dichotomous-context-induced superselection rule for a one-dimensional completed scattering}


\author{N. L. Chuprikov 
}


\institute{N. L. Chuprikov \at
              Tomsk State Pedagogical University, 634041, Tomsk, Russia \\
              \email{chnl@tspu.edu.ru}           
}

\date{Received: date / Accepted: date}

\maketitle

\begin{abstract}
It is shown that in the state space of a quantum particle involved in a one-dimensional completed scattering (OCS) there is a superselection rule
(SSR) induced by the dichotomous physical context that determines the properties of a quantum ensemble of scattering particles in the disjoint
spatial regions located on the different sides of the potential barrier at asymptotically large distances from it. At the initial stage of the OCS
this context is associated, in the general case, with two sources of particles located on the different sides of the barrier; and at the final
stage it is associated, in the general case, with two particle detectors located on different sides of the barrier. The role of a superselection
operator is played by the Pauli matrix $\sigma_3$ that divides the space of in- and out-asymptotes into two coherent sectors. In the course of the
OCS, its unitary quantum dynamics crosses these sectors. In particular, in the scattering problem with one source and two detectors, a pure
initial state of a particle is converted into a mixed final state. According to this rule, the OCS is a complex quantum process consisting of two
coherent subprocesses -- transmission and reflection; the average value of any observable as well as characteristic times can be defined for the
subprocesses only. It is shown that the quantum dynamics of both subprocesses at all stages of scattering is uniquely determined by the initial
state of the whole ensemble of particles and by the final states of transmitted and reflected particles.

\keywords{superselection rule \and one-dimensional completed scattering \and coherent superposition of macroscopically distinct states}
\end{abstract}

\newcommand{\ppp}{\mbox{\hspace{5mm}}}
\newcommand{\ooo}{\mbox{\hspace{3mm}}}
\newcommand{\ooa}{\mbox{\hspace{1mm}}}
\newcommand{\ppd}{\mbox{\hspace{18mm}}}
\newcommand{\ppt}{\mbox{\hspace{34mm}}}
\newcommand{\ppo}{\mbox{\hspace{10mm}}}

\section{Introduction}

The purpose of this article is to introduce a new quantum mechanical model of scattering a particle on a one-dimensional potential barrier which
is nonzero in a limited spatial interval. It is assumed that at the initial stage of scattering a (pure) state of a particle is described by a
sufficiently narrow in the momentum space wave packet which is to the left of the barrier at a distance substantially exceeding the packet's
width. At the next stage this wave packet interacts with the potential barrier, whereby it splits into two parts and, as a result, at the final
stage the state of a particle represents the superposition of the transmitted and reflected wave packets moving in the non-overlapping spatial
regions lying on different sides of the barrier. Such a process (providing that spreading of the transmitted and the reflected wave packets is
sufficiently slow in the course of the OCS; otherwise, splitting the initial wave packet into the transmitted and reflected components does not
occur) is called a one-dimensional {\it completed} scattering (OCS).

It can be assumed that the very name and purpose of the article can cause confusion among most of the experts on quantum theory since the model of
scattering a quantum particle on potential barriers of a simple form -- a potential step and rectangular potential barrier -- are included in many
textbooks on quantum mechanics as examples of an exhaustive description of single-particle quantum scattering processes. In such situation, any
revision of the conventional model of the OCS is perceived as an encroachment on the foundations of quantum mechanics, what is indeed
unacceptable, since this theory underlies all of modern physics.

In addition, the very idea that the OCS can be associated with one or another superselection rule (SSR), contradicts the existing
quantum-mechanical practice. At the present the SSRs are associated with the prohibition of coherent superpositions of pure states of particles
with different spins, electric charges, masses, etc. (see \cite{Wick,Heg} and \cite{Par1,Par2,Par3,Hor1,Hor2,Hor3,Ear,Sch}). With regard to pure
one-particle states associated with the same spin, charge and mass, the modern quantum theory does not imply the existence of SSRs which would
restrict the action of the superposition principle in this class of states.

In this regard, it is important to dwell in detail on the motives which, for all that, force us to delete the problems of scattering a particle on
the potential step and rectangular potential barrier from the category of already solved problems and put them on a par with such fundamental
problems of quantum mechanics as the double-slit experiment and Schrodinger's cat paradox.

\section{The OCS as the problem of adequate description of microcat-states} \label{micro}

As is known, in the formulation of the problem for the OCS the source of particles can be placed both to the left of the barrier and to the right
of it. Thus, the Hilbert space consisting of pure states of a particle involved in the OCS must contain both those states that describe a particle
when it impinges the barrier from the left and those states that describe the particle when it impinges the barrier from the right. Moreover,
according to the modern formulation of the superposition principle, any coherent superposition of these two types of pure states (the problem with
the bilateral incidence of a particle on the barrier (see, e.g., \cite{Mad,Nus})) represents a new pure state that also belongs to this space.

In order to stress a paradoxical character of this requirement, let us consider the problem with two sources of particles in the following
formulation. Let the potential barrier of width $d=2a$ be nonzero in the interval $[-a,a]$, and the left and right sources be located in the
spatial regions $\cal{A}$ and $\cal{B}$, respectively, far enough from the barrier boundaries. And let $|\Psi_{\cal{A}}\rangle$ and
$|\Psi_{\cal{B}}\rangle$ be the (normalized by unit) non-stationary states of the ensembles of particles emitted by the left and right source,
respectively. At $t=0$ in the laboratory reference frame, the maximums of the wave packets $\Psi_{\cal{A}}(x,t)$ and $\Psi_{\cal{B}}(x,t)$ (with
the equal width $l$) are located at the points $-L$ and $+L$, respectively; $L\gg l\gg d$. The corresponding average values of the momentum
operator $\hat{P}$ are $+\hbar k_0$ and $-\hbar k_0$. Then, in the problem with two sources (in each single experiment only one of the two sources
is triggered at the moment $t=0$), the state of the ensemble of particles can be written in the form
$|\Psi_{\cal{AB}}\rangle=(e^{i\phi_{\cal{A}}}|\Psi_{\cal{A}}\rangle+e^{i\phi_{\cal{B}}}|\Psi_{\cal{B}}\rangle)/\sqrt{2}$, where $\phi_{\cal{A}}$
and $\phi_{\cal{B}}$ are real phases.

According to the above requirement the state $|\Psi_{\cal{AB}}\rangle$ must be considered as a pure one and, hence, the corresponding quantum
ensemble of particles cannot be divided into parts, in principle. And this applies to all instants of time $t$, including the instants $t>0$
immediately after triggering one of these two particle sources. But this means that, according to this requirement, even at these instants of time
a particle is simultaneously located {\it both} in the region $\cal{A}$ {\it and} in the region $\cal{B}$, and, for this reason, it 'knows' about
the existence of both phases $\phi_{\cal{A}}$ and $\phi_{\cal{B}}$, which determine its dynamics at the later stages of scattering.

But simultaneous presence of a particle in two disjoint spatial regions located at a macroscopically large distance from each other contradicts
relativity theory. Moreover, this requirement contradicts the fundamental tenets of probability theory, according to which the states
$|\Psi_{\cal{A}}\rangle$ and $|\Psi_{\cal{B}}\rangle$ associated with different physical conditions (contexts), describe {\it different}
statistical ensembles of particles. As it is stressed in \cite{Khr}, "Two collectives of particles moving under two macroscopically distinct
contexts form two different statistical ensembles"; {\it "probabilistic data generated by a few collectives\ldots cannot be described by a single
Kolmogorov space" (ibid)} (see also \cite{Acc}). Thus, according to probability theory, the superposition $|\Psi_{\cal{AB}}\rangle$ must be
treated as a mixed state.

Exactly the same situation arises at the final stage of the OCS, when the state of a particle represents a coherent superposition of the
transmitted and reflected wave packets; the former moves towards the detector located in the region $\cal{B}$ and the latter moves towards the
detector located in the region $\cal{A}$. Again, if we considered this superposition as a pure state, we would inevitably came to the conclusion
that a particle just before it was detected by one of the detectors, was both in the region $\cal{A}$ and in the region $\cal{B}$, simultaneously.

So, at the initial and/or final stages of the OCS the particle states represent {\it coherent superpositions of macroscopically distinct states}.
In the literature on fundamental questions of quantum mechanics (see, e.g., \cite{Leg}) superpositions of such a kind are referred to as
"microcat-states". For most physicists this notion is associated, first of all, with the states of a particle in the double-slit experiment as
well as with those of a radioactive atom in the Shcr\"{o}dinger cat paradox. Generally, namely these two examples are used for illustrating the
paradoxical properties of micro-objects when they are in microcat-states: in the double-slit experiment the particle can simultaneously pass
through two slits in the screen; a radioactive atom, after a time equal to its half-life, can be decayed and undecayed simultaneously.

Due to these paradoxical properties, the double-slit experiment (the only mystery of quantum mechanics, by words of Richard Feynman) and the
Shcr\"{o}dinger's cat paradox (which is usually interpreted as the measurement problem) for many years have been at the center of endless debate
about the fundamental problems of interpreting the wave function and quantum mechanics. And so far none of these problems has received a
universally acceptable solution. Researchers are forced to return again and again to the mysterious two-slit experiment, to the measurement
problem and to the problem of interpretation of quantum theory.

The so-called tunneling time problem associated with the temporal aspects of OCS has been a stumbling block for researchers for a long time. After
the discovery of the tunneling phenomenon, reviews on this problem have regularly appeared right up to 2006 year, when the last review article
\cite{Win}  has been published. As far as we know, no one new review on this topic has appeared after this review, and, at first glance, this fact
would mean that the problem of tunneling time has been resolved. However, the critical analysis in \cite{Win} of existing definitions of the
tunneling time as well as the polemics with opponents, which is presented by the author at the end of his article, say otherwise.

That is, in reality, the standard quantum mechanical model of the OCS does not give a proper description of this process. Moreover, we consider
that the problem of description of this (seemingly simple) quantum process stands on a par with the above mentioned fundamental problems of
quantum theory. And the key to solving all these problems is the same. Our goal is to show that the Schr\"{o}dinger's cat paradox arises precisely
at the micro-level, rather than somewhere on the way from the microcosm to the macrocosm, as is commonly believed (see, e.g., \cite{Zur,Bub}).
This paradox should be considered as the problem of a proper quantum mechanical description of microcat-states, rather than as the measurement
problem. We believe that modern quantum theory of microcat-states distorts the true nature of such states and namely this reason makes it
impossible a consistent interpretation of quantum phenomena where such states emerge and quantum theory itself (and the ensemble interpretation
\cite{Bal,Bal1,Hom} is no exception).

A principal error in the current theory of microcat-states is that the superposition principle of classical physics (where waves and wave packets
(electromagnetic, sound, etc.) can simultaneously pass through two slits in the screen, can split into parts in the course of scattering as well
as can move, after scattering, simultaneously in several non-overlapping spatial regions) is applied {\it without any reservations} to
quantum-mechanical 'probability waves'. Thus, it is completely ignored the fact that in the one-particle quantum theory the probability wave
describes ultimately the {\it particle} dynamics, rather than the {\it wave} dynamics. As a result, it is violated the fundamental principle
according to which a particle cannot simultaneously pass through two slits in the screen, cannot move in the disjoint spatial regions that lie on
different sides of the potential barrier, etc. At the same time this principle must not be considered in the nonrelativistic quantum theory as
less important than the superposition principle. Thus, in the class of microcat-states these two principles must be reconciled with each other.

Our goal is to show, by the example of the OCS, that the Hilbert space, associated with any one-particle quantum process in which there appear
microcat-states, has a nontrivial structure. In this space there is a superselection rule according to which microcat-states must be treated as
mixed states. Only with this interpretation of microcat-states quantum mechanics is really becoming a universal theory.

\section{Stationary states of a particle in the formalism of the transfer and scattering matrices.} \label{station}

Let us consider a spinless non-relativistic particle which is scattered by the potential barrier $V(x)$ given in the spatial interval $[-a,a]$. We
begin our analysis of this process with solving the stationary Schr\"{o}dinger equation for a particle with energy $E=(\hbar k)^2/2m$; $m$ is the
particle's mass. In the general case the wave function $\Psi(x;k)$, beyond the interval $[-a,a]$, can be written in the form
\begin{eqnarray} \label{1}
\Psi(x;k)=\left\{
\begin{array}{cc}
A_{L,in}(k)\ooa e^{ikx}+A_{L,out}(k)\ooa e^{-ikx},\ppp x\leq -a; \\
A_{R,out}(k)\ooa e^{ikx}+A_{R,in}(k)\ooa e^{-ikx},\ppp x\geq +a
\end{array}\right.
\end{eqnarray}
Its amplitudes in the regions $x\leq -a$ and $x\geq a$ are linked by the transfer matrix $\textbf{Y}(k)$:
\begin{eqnarray} \label{111}
\left(\begin{array}{cc} A_{L,in} \\ A_{L,out} \end{array} \right)=\textbf{Y} \left(\begin{array}{cc} A_{R,out} \\ A_{R,in}
\end{array} \right);\ppp \textbf{Y}=\left(
\begin{array}{cc}
q & p \\
p^* & q^*
\end{array} \right);
\end{eqnarray}
where $q(-k)=q^*(k)$, $p(-k)=p^*(k)$. According to \cite{Ch8}, for any potential barrier given in the interval $[x_1,x_2]$ the transfer-matrix
elements can be written as follows,
\begin{eqnarray} \label{2}
q=\frac{1}{\sqrt{T(k)}}\ooa e^{i\left[k(x_2-x_1)-J(k)\right]},\ppp p=i\sqrt{\frac{R(k)}{T(k)}}\ooa
e^{i[-k(x_2+x_1)+F(k)]},\ppp R=1-T;
\end{eqnarray}
$T(-k)=T(k)$, $J(-k)=-J(k)$, $F(-k)=\pi-F(k)$; for the case considered $x_2-x_1=d$ and $x_2+x_1=0$. For any symmetric potential barrier,
$V(-x)=V(x)$, the phase $F$ takes only two values: $0$ or $\pi$. In this case, a piecewise-constant function $F(k)$ has discontinuities at the
points where the reflection coefficient equals to zero.

Note that the scattering parameters (the transmission $T$ and reflection $R$ coefficients, as well as the phases $J$ and $F$) can be calculated
(analytically or numerically) for potential barriers of any form. For this purpose one can use either analytical expressions in \cite{Ch8}, if
$V(x)$ is the rectangular potential barrier or the $\delta$-potential, or recurrence relations, if $V(x)$ represents a system of
$\delta$-potentials and piecewise continuous potential barriers. Thus, we can further assume that the matrix $\textbf{Y}(k)$ is known.

We will also assume that the scattering matrices that link the amplitudes $A_{L,out}$ and $A_{R,out}$ of outgoing waves with the amplitudes
$A_{L,in}$ and $A_{R,in}$ of incoming waves are known too. This linkage can be written by two ways. And since both variants will be important for
our approach (see Section \ref{colomn}) we introduce two scattering matrices -- $\textbf{S}_k$ and $\textbf{S}_x$:
\begin{eqnarray} \label{3}
\left(\begin{array}{cc} A_{R,out} \\ A_{L,out} \end{array} \right)=\textbf{S}_k \left(\begin{array}{cc} A_{L,in} \\
A_{R,in}
\end{array} \right),\ooa \textbf{S}_k=\frac{1}{q}\left(
\begin{array}{cc}
1 & -p \\
p^* & 1
\end{array} \right);\ooo
\left(\begin{array}{cc} A_{L,out} \\ A_{R,out} \end{array} \right)=\textbf{S}_x \left(\begin{array}{cc} A_{L,in} \\
A_{R,in}
\end{array} \right),\ooa \textbf{S}_x=\frac{1}{q}\left(
\begin{array}{cc}
p^* & 1 \\
1 & -p
\end{array} \right)
\end{eqnarray}
It is assumed that, among these four amplitudes, the ones $A_{L,in}(k)$ and $A_{R,in}(k)$ are independent; they are determined in the region $k>0$
and obey the condition $|A_{L,in}(k)|^2+|A_{R,in}(k)|^2=1$. (When changing the sign of the wave number $k$ incoming and outgoing waves swap roles,
due to the time reversibility of quantum mechanics: $A_{L,in}(-k)\equiv A_{L,out}^\prime(k)$, $A_{R,in}(-k)\equiv A_{R,out}^\prime(k)$,
$A_{L,out}(-k)\equiv A_{L,in}^\prime(k)$, $A_{R,out}(-k)\equiv A_{R,in}^\prime(k)$. In this case the new (primed) variables, as the old
amplitudes, are linked by the same relationship (\ref{3}).

Our next step is to determine the space, built of the in - and out-asymptotes of the OCS -- wave packets that describe non-stationary localized
states of a particle at the initial and final stages of the OCS, respectively. In this case, we must take into account that the OCS implies the
lack of any interaction between the left and right asymptotes at these stages of scattering.

\section{The space of in- and out-asymptotes of the OCS}

Considering that $A_{L,out}$ and $A_{R,in}$ are the amplitudes of waves that move on the $OX$-axis from the right to the left and that
$E(-k)=E(k)$, in- and out-asymptotes of the OCS, in the $k$-representation, can be written in the form:
$$\Psi_{L,in}(k,t)=A_{L,in}(k)e^{-iE(k)t/\hbar},\ooo \Psi_{R,in}(k,t)=A_{R,in}(-k)e^{-iE(k)t/\hbar}$$ are in-asymptotes (wave packets), which are
localized in the regions $\cal{A}$ and $\cal{B}$ (see Section \ref{micro}), respectively, and which move toward the barrier;
$$\Psi_{L,out}(k,t)=A_{L,out}(-k)e^{-iE(k)t/\hbar},\ooo \Psi_{R,out}(k,t)=A_{R,out}(k)e^{-iE(k)t/\hbar}$$ are
wave packets which are localized in the regions $\cal{A}$ and $\cal{B}$, respectively, and which move away from the barrier. It should be stressed
that the in- and out-asymptotes describe different stages of the OCS; so that the variable $t$ takes different values for these asymptotes.

The left and right components of these two asymptotes must meet the following conditions: firstly, they must be non-zero in the different regions,
both in the $x$-space and in the $k$-space,
\begin{eqnarray} \label{9}
\langle\Psi_{L,in}|\Psi_{R,in}\rangle=\langle\Psi_{L,out}|\Psi_{R,out}\rangle=0;
\end{eqnarray}
secondly, these functions must ensure the existence of average values of all finite degrees of the operators of the coordinate $\hat{X}$ and
momentum $\hat{P}$ of a particle.

To ensure the fulfillment of these requirements in the $k$-space, we will assume that the independent amplitudes $A_{L,in}(k)$ and $A_{R,in}(k)$
belong to the spaces ${\cal{S}}(0,\infty)$ and ${\cal{S}}(-\infty,0)$, respectively; where ${\cal{S}}(0,\infty)$ is the Schwartz subspace that
consists of infinitely differentiable functions which are zero on the semiaxis $(-\infty,0)$ and diminish in the limit $k\to\infty$ more rapidly
than any power function; ${\cal{S}}(-\infty,0)$ is the Schwartz subspace too, but it consists of functions which are zero on the semiaxis
$(0,\infty)$. Thus, the asymptotes $\Psi_{L,in}(k,t)$ and $\Psi_{R,out}(k,t)$ belong to the subspace ${\cal{S}}(0,\infty)$, and the ones
$\Psi_{L,out}(k,t)$, but $\Psi_{R,in}(k,t)$ belong to the subspace ${\cal{S}}(-\infty,0)$. As a consequence, in the $k$-representation the space
of asymptotes represents the sum of two disjoint subspaces ${\cal{S}}(-\infty,0)$ and ${\cal{S}}(0,\infty)$.

In addition, we assume that the asymptotes $\Psi_{L,in}$ and $\Psi_{R,out}$ satisfy these requirements also in the $x$-space. This can always be
achieved with considering the shift theorem in the theory of Fourier transforms. For example, if the tentative expressions for the amplitudes
$A_{L,in}(k)$ and $A_{R,in}(k)$ were such that $\langle\Psi_{L,in}|\hat{X}|\Psi_{L,in}\rangle=\langle\Psi_{R,in}|\hat{X}|\Psi_{R,in}\rangle=0$
what is unacceptable. Then the maximums of the wave packets with the amplitudes $A_{L,in}(k)e^{ikL}$ and $A_{R,in}(k)e^{-ikL}$, where the length
$L$ is much larger then the width of wave packets, will already be positioned at the points $x=-L$ and $x=L$, that is, far enough from the
barrier.

Thus, in the selected spaces of functions the in-asymptotes $\Psi_{L,in}$ and $\Psi_{R,in}$, as well as the out-asymptotes $\Psi_{L,out}$ and
$\Psi_{R,out}$, do not overlap each other both in the $k$-space and in the $x$-space. On the $OX$-axis the left components of both asymptotes are
localized in the interval $(-\infty,-a)$, and the right components are in the interval $(a,\infty)$. Thus, for the subspaces to which these
components belong more detailed notations are needed. Namely, we will assume further that in the $k$- and $x$-representations
\begin{eqnarray*}
\Psi_{L,in}(k,t),\ooa \Psi_{R,out}(k.t)\in {\cal{S}}(\Omega_k^+),\ooo \Psi_{R,in}(k,t),\ooa\Psi_{L,out}(k,t)\in {\cal{S}}(\Omega_k^-),\\
\Psi_{R,in}(x,t),\ooa\Psi_{R,out}(x,t)\in {\cal{S}}(\Omega_x^+),\ooo \Psi_{L,in}(x,t),\ooa \Psi_{L,out}(x.t)\in {\cal{S}}(\Omega_x^-);
\end{eqnarray*}
where $\Omega_k^+$ and $\Omega_k^-$ are, respectively, the semiaxes $(0,\infty)$ and $(-\infty,0)$ in the $k$-space; while $\Omega_x^+$ and
$\Omega_x^-$ are, respectively, the intervals $(a,\infty)$ and $(-\infty,-a)$ in the $x$-space.

According to the modern quantum theory of the OCS \cite{Rob,Mad,Madr}, the states of a particle involved in the OCS form the rigged (equipped)
Hilbert space ${\cal{H}}^{rig}$ -- a Gelfand triplet $\Phi \subset {\cal{H}} \subset \Phi^\times$, where ${\cal{H}}$ is a Hilbert space; $\Phi$ is
the space of 'physical states'; $\Phi^\times $ is the space of antilinear functionals over $\Phi$, which includes right eigenvectors of
one-particle operators $\hat{X}$ and $\hat{P}$ (the corresponding {\it bra}-vectors belong to the space $\Phi^\prime$ of linear functionals over
$\Phi$).

The term 'physical states' implies that for such states expectation (average) values exist for any finite degree of the operators $\hat{X}$ and
$\hat{P}$, and, by \cite{Rob,Mad,Madr}, such states belong to the Schwartz space ${\cal{S}}$ which is invariant with respect to the
Fourier-transform. This means, in particular, that asymptotes which describe physical states of a particle at the initial and final stages of the
OCS belong to the Schwartz space too.

However, by our approach the space of asymptotes $\Phi\equiv\Phi_{as}$ has a more complex structure. If one considers asymptotes in the
$k$-representation, then $\Phi_{as}={\cal{S}}(\Omega_k^-)\oplus{\cal{S}}(\Omega_k^+)$. In the $x$-representation
$\Phi_{as}={\cal{S}}(\Omega_x^-)\oplus{\cal{S}}(\Omega_x^+)$. Thus, there are reasons to believe that, in the case of the OCS, the space
${\cal{H}}^{rig}$ has, too, a more complex structure than it was previously assumed.

\section{In- and out-asymptotes of the OCS as two-component wave functions} \label{colomn}

Since the left and right components of the in-asymptote $|\Psi_{in}\rangle=|\Psi_{L,in}\rangle+|\Psi_{R,in}\rangle$ and
out-asymptote $|\Psi_{out}\rangle=|\Psi_{L,out}\rangle+|\Psi_{R,out}\rangle$ belong to the disjoint spaces, their scalar
products equal to zero and the expressions for the norms of the vectors $|\Psi_{in}\rangle$ and $|\Psi_{out}\rangle$ do
not contain interference terms. That is,
$\langle\Psi_{in}|\Psi_{in}\rangle=\langle\Psi_{L,in}|\Psi_{L,in}\rangle+\langle\Psi_{R,in}|\Psi_{R,in}\rangle=1$,
$\langle\Psi_{out}|\Psi_{out}\rangle=\langle\Psi_{L,out}|\Psi_{L,out}\rangle+\langle\Psi_{R,out}|\Psi_{R,out}\rangle=1$.

The scattering matrix formalism prompts us that the in- and out-asymptotes of the OCS --  two-component wave functions --
can be presented, similarly to the Pauli spinor, in the form of two-component columns. Thus we will believe further that
any two asymptotes $|\chi\rangle$ and $|\psi\rangle$ can be written as $\left(\begin{array}{cc} \chi_1
\\ \chi_2
\end{array} \right)$ and $\left(\begin{array}{cc} \psi_1
\\ \psi_2 \end{array} \right)$, respectively, and their norms and scalar product are defined by the expressions,
$\langle\chi|\chi\rangle=\langle\chi_1|\chi_1\rangle+\langle\chi_2|\chi_2\rangle$,
$\langle\psi|\psi\rangle=\langle\psi_1|\psi_1\rangle+\langle\psi_2|\psi_2\rangle$,
$\langle\chi|\psi\rangle=\langle\chi_1|\psi_1\rangle+\langle\chi_2|\psi_2\rangle$.

Note, the conformity between the components of an asymptote and corresponding column depends on the formalism of the
scattering matrix which is taken as the basis for the transition to columns. As it turns out, when we consider these
asymptotes in the $k$-representation then we have to use the formalism of the scattering matrix $\textbf{S}_k$; while in
the $x$-representation we have to use the formalism of the scattering matrix $\textbf{S}_x$. And of importance is to
stress that in this case we can use {\it either} the $k$-representation {\it or} the $x$-representation.

\subsection{$k$-representation}

Note that the matrix $\textbf{S}_k$ acts in the space of columns whose first elements describe waves moving along the
$OX$-axis from the left to the right, while the second elements describe waves moving in the opposite direction. In other
words, the first elements of such columns are functions to belong to the subspace ${\cal{S}}(\Omega_k^+)$, while the
second ones are functions that belong to the subspace ${\cal{S}}(\Omega_k^-)$. In this case the asymptotes can be
rewritten in the form
$|\Psi_{in}\rangle=\left(\begin{array}{cc} \Psi_{L,in} \\
\Psi_{R,in}
\end{array} \right)$, $|\Psi_{out}\rangle=\left(\begin{array}{cc} \Psi_{R,out}
\\ \Psi_{L,out} \end{array} \right)$. The corresponding bra-vectors represent rows: $\langle\Psi_{in}|=\left(\Psi^*_{L,in}, \Psi^*_{R,in}\right)$,
$\langle\Psi_{out}|=\left(\Psi^*_{R,out}, \Psi^*_{L,out} \right)$.

Let us consider such a pair of vectors $|\phi_{k'}^{(1)}\rangle$ and $|\phi_{k'}^{(2)}\rangle$ with the parameter $k'>0$, as well as such a pair
of vectors $|\phi_{x'}^{(1)}\rangle$ and $|\phi_{x'}^{(2)}\rangle$ with the parameter $x'$, that
\begin{eqnarray*}
\phi_{k'}^{(1)}(k)=\left(\begin{array}{cc} \delta(k-k')
\\ 0 \end{array} \right),\ooo \phi_{k'}^{(2)}(k)=\left(\begin{array}{cc} 0 \\ \delta(k+k')
\end{array} \right);\ooo \phi_{x'}^{(1)}(k)=\left(\begin{array}{cc} e^{-ikx'}
\\ 0 \end{array} \right),\ooo \phi_{x'}^{(2)}(k)=\left(\begin{array}{cc} 0 \\ e^{ikx'} \end{array} \right).
\end{eqnarray*}
Is is evident that the first pair of these vectors gives eigenvectors of the momentum operator $\hat{P}=\hbar k$; with
$\hat{P}|\phi_{k'}^{(1)}\rangle=+\hbar k'|\phi_{k'}^{(1)}\rangle$, $\hat{P}|\phi_{k'}^{(2)}\rangle=-\hbar
k'|\phi_{k'}^{(2)}\rangle$, and $\langle\phi_{k'}^{(1)}|\phi_{k'}^{(2)}\rangle=0$. The second pair gives eigenvectors of
the position operator $\hat{X}=i\frac{d}{dk}$; with $\hat{X}|\phi_{x'}^{(1)}\rangle=+x'|\phi_{x'}^{(1)}\rangle$,
$\hat{X}|\phi_{x'}^{(2)}\rangle=-x'|\phi_{x'}^{(2)}\rangle$, and $\langle\phi_{x'}^{(1)}|\phi_{x'}^{(2)}\rangle=0$.

The stationary in- and out-asymptotes with a given wave-number $k'$ can be written now in the form
$\Psi_{in}(k;k')=A_{L,in}(k)\phi_{k'}^{(1)}(k)+A_{R,in}(k)\phi_{k'}^{(2)}(k)$ and
$\Psi_{out}(k;k')=A_{R,out}(k)\phi_{k'}^{(1)}(k)+A_{L,out}(k)\phi_{k'}^{(2)}(k)$.

\subsection{$x$-representation}

In the $x$-representation we have to use the formalism of the scattering matrix $\textbf{S}_x$. This matrix acts in the
space of columns whose first elements describe waves moving the $OX$-axis to the left of the barrier, while their second
elements describe waves that move to the right of the barrier. Now, the first elements of columns are functions that
belong to the subspace ${\cal{S}}(\Omega_x^-)$, while the second elements are functions that belong to the subspace
${\cal{S}}(\Omega_x^+)$. Thus, now
$|\Psi_{in}\rangle=\left(\begin{array}{cc} \Psi_{L,in} \\
\Psi_{R,in}
\end{array} \right)$ è $|\Psi_{out}\rangle=\left(\begin{array}{cc} \Psi_{L,out}
\\ \Psi_{R,out} \end{array} \right)$.

Let us now consider such a pair of vectors $|\chi_{k'}^{(1)}\rangle$ and $|\chi_{k'}^{(2)}\rangle$ with the parameter
$k'$, as well as such a pair of vectors $|\chi_{x'}^{(1)}\rangle$ and $|\chi_{x'}^{(2)}\rangle$ with the parameter $x'>0$,
that
\begin{eqnarray*}
\chi_{k'}^{(1)}(x)=\left(\begin{array}{cc} e^{ik'x}
\\ 0 \end{array} \right),\ooo \chi_{k'}^{(2)}(x)=\left(\begin{array}{cc} 0 \\ e^{-ik'x} \end{array} \right);\ooo
\chi_{x'}^{(1)}(x)=\left(\begin{array}{cc} \delta(x+x')
\\ 0 \end{array} \right),\ooo \chi_{x'}^{(2)}(x)=\left(\begin{array}{cc} 0 \\ \delta(x-x')
\end{array} \right).
\end{eqnarray*}
It is evident that the first pair of vectors gives eigenvectors of the momentum operator $\hat{P}=-i\hbar \frac{d}{dx}$; with
$\hat{P}|\chi_{k'}^{(1)}\rangle=+\hbar k'|\chi_{k'}^{(1)}\rangle$, $\hat{P}|\chi_{k'}^{(2)}\rangle=-\hbar k'|\chi_{k'}^{(2)}\rangle$, and
$\langle\chi_{k'}^{(1)}|\chi_{k'}^{(2)}\rangle=0$. The second pair gives eigenvectors of the position operator $\hat{X}=x$:
$\hat{X}|\chi_{x'}^{(1)}\rangle=-x'|\chi_{x'}^{(1)}\rangle$, $\hat{X}|\chi_{x'}^{(2)}\rangle=+x'|\chi_{x'}^{(2)}\rangle$, and
$\langle\chi_{x'}^{(1)}|\chi_{x'}^{(2)}\rangle=0$.

The stationary in- and out-states with a given wave-number $k$ can be written now in the form
$\Psi_{in}(x;k)=A_{L,in}(k)\chi_{k}^{(1)}(x)+A_{R,in}(k)\chi_{k}^{(2)}(x)$ and
$\Psi_{out}(x;k)=A_{L,out}(k)\chi_{k}^{(1)}(x)+A_{R,out}(k)\phi_{k}^{(2)}(x)$.

\section{The Pauli matrix $\sigma_3$ as a superselection operator in the space of the OCS asymptotes}

\subsection{$k$-representation}

Note that the vectors $|\phi_{k'}^{(1)}\rangle$, $|\phi_{k'}^{(2)}\rangle$, $|\phi_{x'}^{(1)}\rangle$ and
$|\phi_{x'}^{(2)}\rangle$ are eigenvectors of the Pauli matrix $\sigma_3=\left(\begin{array}{cc} 1 & 0 \\ 0 & -1
\end{array} \right)$. Indeed,
\begin{eqnarray*}
\sigma_3|\phi_{k'}^{(1)}\rangle=|\phi_{k'}^{(1)}\rangle,\ooo \sigma_3|\phi_{k'}^{(2)}\rangle=-|\phi_{k'}^{(2)}\rangle;\ppp
\sigma_3|\phi_{x'}^{(1)}\rangle=|\phi_{x'}^{(1)}\rangle,\ooo \sigma_3|\phi_{x'}^{(2)}\rangle=-|\phi_{x'}^{(2)}\rangle.
\end{eqnarray*}
Thus, the set of eigenvectors of the operator $\sigma_3$ represents the basis in the space $\Phi^\times_{as}$ of
ket-vectors, of which the asymptotes are built in the $k$-representation. And this space contains also the eigenvectors of
the momentum and position operators. In other words, although $\hat{X}$ and $\hat{P}$ do not commute with each other, each
of them commutes with the operator $\sigma_3$. It is also important to stress (see \cite{Hor2}) that the operator
$\sigma_3$ can be expressed via the projection operators
$P_+= \left(\begin{array}{cc} 1 & 0 \\
0 & 0 \end{array} \right)$ è $P_-= \left(\begin{array}{cc} 0 & 0 \\ 0 & 1 \end{array} \right)$. Namely, $\sigma_3=P_+-P_-$.

Since the state space of a particle  involved in the OCS is complete (see, e.g., \cite{Madr}) (and no matter what stage of
this process is regarded), the self-adjoint operator $\sigma_3$ can be treated (see \cite{Hor2} as well as \cite{Hor3}) as
the superselection operator which divides the state space ${\cal{H}}^{rig}_{as}$, in the $k$-representation, into two
coherent (superselection) sectors (${\cal{H}}^{rig}_{as}$ is the rigged Hilbert space ${\cal{H}}^{rig}$ associated with
the initial and final stages of the OCS):
\begin{eqnarray} \label{12}
{\cal{H}}^{rig}_{as}={\cal{H}}^{rig}_{as}(\Omega_k^+)\oplus {\cal{H}}^{rig}_{as}(\Omega_k^-);
\end{eqnarray}
\begin{eqnarray*}
{\cal{H}}^{rig}_{as}(\Omega_k^+)=\Phi_{as}(\Omega_k^+) \subset L_2^{as}(\Omega_k^+) \subset \Phi_{as}^\times(\Omega_k^+),\ooo
{\cal{H}}^{rig}_{as}(\Omega_k^-)=\Phi_{as}(\Omega_k^-) \subset L_2^{as}(\Omega_k^-) \subset \Phi_{as}^\times(\Omega_k^-).
\end{eqnarray*}
Here the subspace ${\cal{H}}_{as}^{rig}(\Omega_k^+)$ belongs to the coherent sector (let's call it the 'top coherent
sector') that corresponds to the eigenvalue $+1$ of the operator $\sigma_3$, while the subspace
${\cal{H}}_{as}^{rig}(\Omega_k^-)$ belongs to the 'lower coherent sector'  corresponding to the eigenvalue $-1$. It is
evident that $|\phi_{k'}^{(1)}\rangle,\ooa |\phi_{x'}^{(1)}\rangle\in \Phi_{as}^\times(\Omega_k^+)$, and
$|\phi_{k'}^{(2)}\rangle,\ooa |\phi_{x'}^{(2)}\rangle\in \Phi_{as}^\times(\Omega_k^-)$.

According to the modern theory of SSRs \cite{Par1,Par2,Par3,Hor1,Hor2,Hor3} any superposition of pure states from the same
coherent sector represents another pure state in this sector, while any superposition of pure states from different
sectors represents a mixed state. In order to illustrate one feature of such a kind superpositions let us consider the
following example.

Let $\hat{O}$ be a self-adjoint operator and the left and right in-asymptotes $|\Psi_{L,in}\rangle$ and
$|\Psi_{R,in}\rangle$ be the states $|\Psi_{\cal{A}}\rangle$ and $|\Psi_{\cal{B}}\rangle$, respectively (see Section
\ref{micro}); $|\Psi_{\cal{A}}\rangle\in \Phi_{as}(\Omega_k^+)$, and $|\Psi_{\cal{B}}\rangle\in \Phi_{as}(\Omega_k^-)$.
Besides, let $|\psi_\lambda\rangle =|\Psi_{\cal{A}}\rangle+e^{i\lambda}|\Psi_{\cal{B}}\rangle$, $|\psi_\nu\rangle
=|\Psi_{\cal{A}}\rangle+e^{i\nu}|\Psi_{\cal{B}}\rangle$; $\lambda$ and $\nu$ are real phases. Then
\begin{eqnarray*}
\langle\psi_\lambda|\hat{O}|\psi_\lambda\rangle=\langle\psi_\nu|\hat{O}|\psi_\nu\rangle=
\langle\Psi_{\cal{A}}|\hat{O}|\Psi_{\cal{A}}\rangle+\langle\Psi_{\cal{B}}|\hat{O}|\Psi_{\cal{B}}\rangle.
\end{eqnarray*}
That is, at the initial stage of scattering the phases $\lambda$ and $\nu$ are unobservable quantities, what is one of
signs (see \cite{Ear}) that the states $|\psi_\lambda\rangle$ and $|\psi_\nu\rangle$, representing the coherent
superpositions of the left and right asymptotes, are mixed states.

\subsection{$x$-representation}

Since the space of asymptotes given in the $k$-representation splits into two coherent sectors, a similar situation must arise also in the
$x$-representation. Indeed, the vectors $|\chi_{k'}^{(1)}\rangle$, $|\chi_{k'}^{(2)}\rangle$, $|\chi_{x'}^{(1)}\rangle$ and
$|\chi_{x'}^{(2)}\rangle$ are eigenvectors of the matrix $\sigma_3$:
\begin{eqnarray*}
\sigma_3|\chi_{k'}^{(1)}\rangle=|\chi_{k'}^{(1)}\rangle,\ooo \sigma_3|\chi_{k'}^{(2)}\rangle=-|\chi_{k'}^{(2)}\rangle;\ppp
\sigma_3|\chi_{x'}^{(1)}\rangle=|\chi_{x'}^{(1)}\rangle,\ooo \sigma_3|\chi_{x'}^{(2)}\rangle=-|\chi_{x'}^{(2)}\rangle.
\end{eqnarray*}
That is, here, too, two coherent sector arise. The only difference is that, in the $x$-representation, the eigenvalue $+1$
of the superselection operator $\sigma_3$ is associated with functions of the subspace ${\cal{H}}_{as}^{rig}(\Omega_x^-)$
(let's call it the 'left coherent sector', while the eigenvalue $-1$ is associated with functions that belong to the
subspace ${\cal{H}}_{as}^{rig}(\Omega_x^+)$ (let's call it the 'right coherent sector'). Thus, in the $x$-representation,
\begin{eqnarray} \label{13}
{\cal{H}}^{rig}_{as}={\cal{H}}^{rig}_{as}(\Omega_x^-)\oplus {\cal{H}}^{rig}_{as}(\Omega_x^+);
\end{eqnarray}
\begin{eqnarray*}
{\cal{H}}^{rig}_{as}(\Omega_x^-)=\Phi_{as}(\Omega_x^-) \subset L_2^{as}(\Omega_x^-) \subset \Phi_{as}^\times(\Omega_x^-),\ooo
{\cal{H}}^{rig}_{as}(\Omega_x^+)=\Phi_{as}(\Omega_x^+) \subset L_2^{as}(\Omega_x^+) \subset \Phi_{as}^\times(\Omega_x^+).
\end{eqnarray*}
It is evident that $|\chi_{k'}^{(1)}\rangle,\ooa |\chi_{x'}^{(1)}\rangle\in \Phi_{as}^\times(\Omega_x^-)$ and $|\chi_{k'}^{(2)}\rangle,\ooa
|\chi_{x'}^{(2)}\rangle\in \Phi_{as}^\times(\Omega_x^+)$.

\section{The OCS as a coherent superposition of two alternative subprocesses -- transmission (tunneling) and reflection.}

Note that the Hamiltonian $\hat{H}$ does not commute with the superselection operator $\sigma_3$ since the stationary state of a particle cannot
be associated with a single superselection sector. That is, in the case of the OCS the Shcr\"{o}dinger dynamics crosses the superselection
sectors. Let us show this by the example of the scattering problem, when there is only one source of particles and it is located in the region
$\cal{A}$ (see Section \ref{micro}). The corresponding asymptotes are described by the expressions
\begin{eqnarray} \label{14}
\Psi_{in}(x,t)=\int_{-\infty}^\infty A_{L,in}(k)\chi_{k}^{(1)}(x)e^{-iE(k)t/\hbar} dk;\ooo \Psi_{out}(x,t)=\Psi_{L,out}(x,t)+\Psi_{R,out}(x,t),\ppp\\
\Psi_{L,out}=\int_{-\infty}^\infty A_{L,in}(k)\frac{p^*(k)}{q(k)}\chi_{k}^{(1)}(x)e^{-iE(k)t/\hbar} dk,\ooo \Psi_{R,out}=\int_{-\infty}^\infty
A_{L,in}(k)\frac{1}{q(k)}\chi_{k}^{(2)}(x)e^{-iE(k)t/\hbar} dk;\nonumber
\end{eqnarray}
$\langle\Psi_{in}|\Psi_{in}\rangle=\overline{T}+\overline{R}=1$ where $\overline{T}=\langle\Psi_{R,out}|\Psi_{R,out}\rangle$,
$\overline{R}=\langle\Psi_{L,out}|\Psi_{L,out}\rangle$.

In this case the in-asymptote has only the left component that represents a pure state from the subspace $\Phi_{as}(\Omega_x^-)$ -- the left
coherent sector. As regards the out-asymptote, its component $\Psi_{L,out}$ is a pure state that belongs to the subspace $\Phi_{as}(\Omega_x^-)$
-- the left coherent sector; while its component $\Psi_{R,out}$ is a pure state belonging to the subspace $\Phi_{as}(\Omega_x^+)$ -- the right
coherent sector. Thus, the two-component out-asymptote -- a microcat-state --  represents a mixed state. That is, in this scattering problem, the
Shcr\"{o}dinger quantum dynamics crosses the coherent sectors, transforming a pure (initial) state into a mixed (final) state.

Thus, in the case of the OCS the operator $\hat{H}$ cannot be considered, in the traditional quantum mechanical sense, as the operator associated
with a physical observable, because its eigenvectors cannot be associated with some coherent sector. In particular, in the problem (\ref{14}) the
stationary wave function in the $x$-representation can be associated neither with the subspace $\Phi_{as}(\Omega_x^-)$ nor with the subspace
$\Phi_{as}(\Omega_x^+)$.

So, in the quantum theory of the OCS there is a SSR that divides the asymptote space, associated with this process, into two coherent sectors. And
what is important is that this rule does not forbid the two-component asymptotes as allegedly non-physical states. It only forbids to treat them
as pure states. Contrary to the contemporary quantum mechanical model of the OCS this rule forbids to treat this process as a 'pure' quantum
process, indivisible into subprocesses (and, of no importance, whether the unilateral or bilateral incidence of a particle on the barrier is
studied). In particular, the SSR forbids calculation of the average values of physical quantities as well as introduction of characteristic times
for the whole scattering process.

But if so, then in the problem (\ref{14}) with the unilateral incidence of a particle on the barrier an adequate theory of the OCS must treat this
process as the combination of two coherent subprocesses -- transmission (tunneling) and reflection, and only for these subprocesses the average
values of physical quantities as well as characteristic times can be defined. Of course, this implies that the dynamics of each subprocesses is
known at all stages of scattering. At the same time, in the conventional model of the OCS,  the wave functions for these subprocesses are known
only at the final stage of this process. Thus, elaborating a quantum mechanical model of the OCS, consistent with the SSR, faces with the problem
of reconstructing the quantum dynamics of each subprocess at the stages preceding the final stage of scattering. A method of solving this problem
is presented below.

Let $\psi^{tr}(x;k)$ and $\psi^{ref}(x;k)$ are sought wave functions that describe in the problem (\ref{14}) the transmission and reflection
subprocesses, respectively. Then the first two requirements on these functions, which reflect the main specifics of these subprocesses, can be
formulated as follows:
\begin{itemize}
\item[(a)] the superposition of the wave functions of the subprocesses must describe the whole process; that is, $\psi^{tr}(x;k)+\psi^{ref}(x;k)=\Psi(x;k)$
(see (\ref{1}));
\item[(b)] each of these two wave functions must have only one incoming wave and only one outgoing wave; in this case, for $\psi^{tr}(x;k)$ the incoming
wave is to the  left of the barrier and the outgoing one is to the right of the barrier; for $\psi^{ref}(x;k)$ the incoming and outgoing waves lie
to the left of the barrier.
\end{itemize}
Besides, we have also to take into account the fact that for each subprocess the incoming and outgoing waves describe the states of the same
ensemble of free particles. Thus, these states may differ from each other only by a phase factor (see the requirement (c)).

In order to formulate this requirement in the mathematical form we suppose for convenience that in Exp. (\ref{1}) for $\Psi(x;k)$, in the problem
with the unilateral incidence ($A_{R,in}=0$) of a particle on the barrier, the amplitude of the wave incident on the barrier from the left equals
to unit, that is,
\begin{eqnarray} \label{15}
A_{L,in}=1,\ooo A_{R,out}=1/q=\sqrt{T}\exp[i(J-kd)],\ooo A_{L,out}=p^*/q=-i\sqrt{T}\exp[i(J-F-kd)].
\end{eqnarray}
Thus, if $A^{tr}_{L,in}$ and $A^{ref}_{L,in}$ are the amplitudes of incoming waves in $\psi^{tr}(x;k)$ and $\psi^{ref}(x;k)$, respectively, then
the third requirement for the searched-for functions can be written in the form
\begin{itemize}
\item[(c)] $A^{tr}_{L,in}=\frac{1}{q}e^{i\alpha}$ and $A^{ref}_{L,in}=\frac{p^*}{q}e^{i\beta}$; where $\alpha$ and $\beta$ are real phases.
\end{itemize}

But, according to the condition (a), these phases must obey the equation $\frac{1}{q}e^{i\alpha}+\frac{p^*}{q}e^{i\beta}=1$. From here we find
both the phases and the corresponding amplitudes:
\begin{eqnarray} \label{16}
A^{tr}_{L,in}=\sqrt{T}(\sqrt{T}-i\mu\sqrt{R}),\ooo A^{ref}_{L,in}=\sqrt{R}(\sqrt{R}+i\mu\sqrt{T});\ooo \mu=\pm 1.
\end{eqnarray}
It is seen that not only $A^{tr}_{L,in}+A^{ref}_{L,in}=1$, but also $|A^{tr}_{L,in}|^2+|A^{ref}_{L,in}|^2=1$.

Let us now show by the example of symmetric potential barriers that among these two roots only one leads to the sought wave functions
$\psi^{tr}(x;k)$ and $\psi^{ref}(x;k)$. For this purpose let us consider the solution $\Psi^{ref}(x;k)$ of the Shcr\"{o}dinger equation whose
waves in region $x<-a$ have the same amplitudes $A^{ref}_{L,in}$ and $A_{L,out}$ as the function $\psi^{ref}(x;k)$.

According to (\ref{2}), in the region $x>a$ this solution has the amplitudes
\begin{eqnarray} \label{17}
A^{ref}_{R,out}=i\mu\sqrt{R}e^{i(J-kd)},\ooo A^{ref}_{R,in}=-\mu \sqrt{R}(\sqrt{R}+i\mu\sqrt{T})e^{-iF}.
\end{eqnarray}
Thus, with considering Exps. (\ref{15})--(\ref{17}), we obtain
\begin{eqnarray*}
\frac{A^{ref}_{L,in}}{A^{ref}_{R,in}}=\frac{A^{ref}_{R,out}}{A_{L,out}}=-\mu e^{-iF}.
\end{eqnarray*}
Hence, the function $\Psi^{ref}(x;k)$ for $x<-a$ and $x>a$ can be written, respectively, in the form
\begin{eqnarray*}
\Psi^{ref}(x;k)=A^{ref}_{L,in}e^{ikx}+A_{L,out}e^{-ikx},\ooo \Psi^{ref}(x;k)=-\mu
e^{iF}\left(A^{ref}_{L,in}e^{-ikx}+e^{-2iF}A_{L,out}e^{ikx}\right).
\end{eqnarray*}

But, as it was stressed in Section \ref{station}, for symmetrical potential barriers the phase $F$ equals to either zero or $\pi$. Thus, for such
barriers the function $\Psi^{ref}(x;k)$ is odd if $\mu(k)=sign\left[\cos(F(k))\right]$. The function $\Psi^{ref}(x;k)$ corresponding to this root
equals to zero at the origin (that is, at the midpoint of the barrier region) for any value of $k$. This means that particles incident on the
barrier from any side are returned to the initial region without intersecting the midpoint of the barrier region. Hence, since in the considered
problem the source of particles is to the left of the symmetrical barrier, the wave function for the reflection subproces is given by the
expressions
\begin{eqnarray} \label{18}
\psi^{ref}(x;k)\equiv \Psi^{ref}(x;k),\ooo \mbox{if}\ooo x<0;\ppp \psi^{ref}(x;k)\equiv 0,\ooo \mbox{if}\ooo x\geq 0.
\end{eqnarray}
As regards the transmission subprocess, according to the condition (a) $\psi^{tr}(x;k)=\Psi(x;k)-\psi^{ref}(x;k)$. The wave function obtained in
such manner are continuous at the point $x=0$ together with the corresponding probability current density.

(It is interesting that inserting, at the midpoint of the symmetric potential barrier $V(x)$, the fully opaque $\delta$-potential $W\delta(x)$
with the however large power $W$ does not change the symmetry of the problem and, hence, the function $\Psi^{ref}(x;k)$ is a solution to the
scattering problem with two sources of particles, both without and with this $\delta$-potential. But what is important is that now, for the
Shcr\"{o}dinger equation with the added $\delta$-potential, the function $\psi^{ref}(x;k)$ is a solution to describe particles impinging the fully
opaque symmetric barrier from the left. That is, in the modified experiment the function $\psi^{ref}(x;k)$ can be directly observed.)

Note that the root $\mu(k)=-sign\left[\cos(F(k))\right]$ is associated with the even function $\Psi^{ref}(x;k)$ whose first derivative on $x$ is
zero at the point $x=0$ for any value of $k$. That is, in this case, too, particles impinging on the symmetric barrier $V(x)$ from the left and
right do not cross the midpoint $x=0$. However, in this case the rule (\ref{18}) leads to the function $\psi^{ref}(x;k)$ with the discontinuous
probability density at the point $x=0$; what is unacceptable.

Thus, the requirements (a)-(c), together with the continuity requirement, uniquely determine the functions $\psi^{ref}(x;k)$, $\psi^{tr}(x;k)$ and
the corresponding asymptotes
\begin{eqnarray*}
\Psi^{ref}_{in}(x,t)=\int_{-\infty}^\infty A_{L,in}(k)A^{ref}_{L,in}(k)\chi_{k}^{(1)}(x)e^{-iE(k)t/\hbar} dk,\ppp
\Psi^{ref}_{out}(x,t)=\Psi_{L,out}(x,t);
\end{eqnarray*}
\begin{eqnarray*}
\Psi^{tr}_{in}(x,t)=\int_{-\infty}^\infty A_{L,in}(k)A^{tr}_{L,in}(k)\chi_{k}^{(1)}(x)e^{-iE(k)t/\hbar} dk,\ppp
\Psi^{tr}_{out}(x,t)=\Psi_{R,out}(x,t);
\end{eqnarray*}
in this case, with considering (\ref{16}),
$\langle\Psi^{ref}_{in}|\Psi^{ref}_{in}\rangle=\langle\Psi^{ref}_{out}|\Psi^{ref}_{out}\rangle=\overline{R}$ and
$\langle\Psi^{tr}_{in}|\Psi^{tr}_{in}\rangle=\langle\Psi^{tr}_{out}|\Psi^{tr}_{out}\rangle=\overline{T}$. Now we can proceed to the detailed
analysis of the properties of both subprocesses of the OCS.

\section{On peculiarities of the quantum dynamics of the OCS subprocesses}

Thus, while in the $k$-representation the spatial regions associated with different coherent sectors are separated from each other by the point
$k=0$, in the $x$-representation this role, in the case of symmetric potential barriers, is played by the midpoint of the barrier region (in the
considered problem it is the point $x=0$).

For asymmetric barriers the state of affairs remains the same only in the $k$-representation. In the $x$-representation, the point $x_c$ which
restricts the region of motion of reflected particles does not coincide in the general case with the midpoint of the barrier region and depends on
$k$. In this case the sign of $\mu$ is determined, as for symmetric barriers, by the rule $\mu(k)=sign\left[\cos(F(k))\right]$.

The dependence $x_c(k)$ is determined by the shape of an asymmetric barrier. For such barriers, the point $x_c(k)$ can be removed from the
midpoint of the barrier region not more than by $2\pi/k$; in this case the distance $|x_c(k)|$ remains finite in the limit $k\to 0$. Thus, in the
case of an asymmetric potential barrier the values of $x_c(k)$ form the finite interval $[x_c^{min},x_c^{max}]$, the boundaries of which may be
located outside the interval $[-a,a]$. When this happens, the boundaries of the regions $\Omega_x^-$ and $\Omega_x^-$ in which the left and right
asymptotes are localized must be corrected (although, qualitatively, this doesn't change anything).

Note that at the point $x_c(k)$ the functions $\psi^{ref}(x;k)$ and $\psi^{tr}(x;k)$ are continuous together with the corresponding probability
current densities, while their first derivatives on $x$ are discontinuous here. From the viewpoint of the conventional theory of the OCS such
functions have no physical sense, and only solutions of the Shcr\"{o}dinger equation $\Psi(x;k)$ and $\Psi^{ref}(x;k)$, which are everywhere
continuous together with their first derivatives on $x$, describe a particle involved in this scattering process. On the contrary, from the
viewpoint of the found SSR, these are the 'usual' wave functions that have no physical sense. Indeed, the wave function describing the OCS, in
order to be true, must contain comprehensive information about this quantum process. It is evident that the 'usual' wave functions $\Psi(x;k)$ and
$\Psi^{ref}(x;k)$ do not obey this requirement. They do not contain any information about the fact that the quantum dynamics, associated with this
process, traverses the coherent sectors at the point $x_c(k)$.

As regards the wave functions to describe the subprocesses of the OCS, they contain this information. The fact that the quantum dynamics of the
OCS traverses the coherent sectors (either in the $k$-space or in the $x$-space) is reflected in the discontinuous behaviour of the (true) wave
functions $d\psi^{ref}(x;k)/dx$ and $d\psi^{tr}(x;k)/dx$ at the point $x_c$. In this case the incoming waves in the wave functions of both
subprocesses belong to the same coherent sector both in the $k$-space (the sector ${\cal{H}}^{rig}_{as}(\Omega_k^+)$) and in the $x$-space (the
sector ${\cal{H}}^{rig}_{as}(\Omega_x^-)$). Superposition of the corresponding pure in-asymptotes $\Psi^{tr}_{in}$ and $\Psi^{tr}_{in}$ gives the
pure in-asymptote $\Psi_{in}$ which describes the whole process and which lies in the same sector.

In the course of scattering, the subprocesses traverse the coherent sectors either in the $x$-space or in the $k$-space; in particular, the
transmission subprocess transfers the corresponding initial state of a particle from the coherent sector ${\cal{H}}^{rig}_{as}(\Omega_x^-)$ into
the sector ${\cal{H}}^{rig}_{as}(\Omega_x^+)$; while the reflection subprocess transfers the corresponding initial state from the coherent sector
${\cal{H}}^{rig}_{as}(\Omega_k^+)$ into the sector ${\cal{H}}^{rig}_{as}(\Omega_k^-)$. And this entirely agrees with the SSR. Indeed, if the
stationary states $\psi^{ref}(x;k)$ and $\psi^{tr}(x;k)$ were in the same coherent sector in both representations, their superposition would be a
pure state; what contradicts the SSR.

One more important peculiarity of the quantum dynamics of the OCS is that the superposition of the asymptotes $\Psi^{tr}_{in}(x,t)$ and
$\Psi^{ref}_{in}(x,t)$, Indeed, on the one hand, their scalar product
$$\langle\Psi^{tr}_{in}|\Psi^{ref}_{in}\rangle=i\int_{-\infty}^\infty \mu(k)\sqrt{T(k)R(k)}\ooa\big|A_{L,in}(k)\big|^2dk$$ is not zero, because
they belong to the same coherent sector; but on the other hand, since this product is a purely imaginary quantity,
$\langle\Psi^{tr}_{in}|\Psi^{tr}_{in}\rangle+\langle\Psi^{ref}_{in}|\Psi^{ref}_{in}\rangle=1$, what is characteristic to mixed states. In other
words, quantum probabilities that describe the OCS subprocesses behave like classical ones not only at the final stage of scattering, when the
ensembles of transmitted and reflected particles move on the different sides of the barrier, but also at the initial stage, when they move in the
same spatial region located to the right of the barrier.

Note that the norms of the wave packets $\Psi^{tr}(x,t)$ and $\Psi^{ref}(x,t)$ built, respectively, from the stationary solutions $\psi^{tr}(x;k)$
and $\psi^{ref}(x;k)$ do not conserved at the very stage of scattering. This is explained by the fact that the continuity of probability current
density at the point $x_c(k)$ for each $k$ does not guarantee conserving the number of particles in the quantum ensemble which is described by
wave packets. This is so because the probability current density is expressed nonlinearly through the wave function and its first derivative.
(Exception is the wave packet $\Psi^{ref}(x,t)$ when it describes the reflection subprocess in the case of symmetric potential barriers. For such
barriers $x_c^{min}=x_c^{max}=0$ and hence $\psi^{ref}(0;k)=0$ for all values of $k$. Note that the norm of the wave packet $\Psi^{tr}(x,t)$ is
not constant even for such barriers.) That is, at the very stage of scattering $\overline{T}+\overline{R}\neq 1$.

\section{Conclusion}

It is shown that in the state space associated with a particle involved in the OCS there is a SSR induced by the dichotomous physical context that
determines the quantum dynamics of the particle in the spatial regions located on different sides of the potential barrier, at the asymptotically
large distances from it. The role of the superselection operator is played by the Pauli matrix $\sigma_3$ that divides the space of in- and
out-asymptotes of the OCS into two superselection sectors. In the course of the OCS, the quantum dynamics crosses these sectors. In this case the
SSR does not forbid treating this process as a physical process. Rather it forbids treating it as a (pure) quantum process, which is indivisible
into subprocesses. According to this rule, the OCS is a (mixed) quantum process consisting of two coherent subprocesses -- transmission and
reflection. Average values of all physical observables as well as characteristic times can be defined for the subprocesses only.

The wave functions that describe subprocesses at all stages of scattering are uniquely determined by the initial state of the whole ensemble of
particle and the final states of the subensembles of transmitted and reflected particles. On this basis the following peculiarities of the quantum
dynamics of the subprocesses have been revealed: first, at the asymptotically large distances from the barrier, that is, at the initial and final
stages of scattering, the quantum dynamics of the subprocesses is unitary, and the corresponding quantum probabilities behave as classical ones
that describe mutually exclusive random events (subprocesses); second, at the very stage of scattering the unitary character of the quantum
dynamics of each subprocess is violated at the general case, and the corresponding quantum probabilities violate the requirements imposed by
classical probability theory for alternative random events (subprocesses).

The presented model gives reasons to believe that SSRs must act in the quantum mechanical models of all quantum processes in which microcat-states
appear. This means that all existing quantum mechanical models of such processes must be revised with aim of finding the nontrivial structure of
the corresponding Hilbert spaces and SSRs which should act in these spaces. From our point of view, this is the only way that leads to solving
those paradoxes that arise in the framework of the contemporary models of these quantum phenomena. In particular, a logically consistent solving
of the tunneling time problem is possible only on the basis of the model of the OCS, which is presented here.

\end{document}